\documentclass[doublecol]{epl2}

\title{Event-by-event simulation of double-slit experiments with single photons}
\shorttitle{Event-based simulation of double-slit experiments with single photons} 

\author{F. Jin\inst{1} \and S. Yuan\inst{1} \and H. De Raedt\inst{1}\thanks{E-mail: h.a.de.raedt@rug.nl} 
\and K. Michielsen\inst{2}}
\shortauthor{F. Jin \etal}

\institute{
  \inst{1}Department of Applied Physics, Zernike Institute for Advanced Materials,
University of Groningen, Nijenborgh 4, NL-9747 AG Groningen, The Netherlands\\
  \inst{2} EMBD, Vlasakker 21, B-2160 Wommelgem, Belgium
}

\abstract{We present a computer simulation model that reproduces, event-by-event, the
wave mechanical results of double-slit and two-beam interference experiments.
The same model also simulates a one-to-one copy of a single-photon interference experiment with a Fresnel biprism
(Jacques V. {\sl et al.}, {\sl Eur. Phys. J. D}, {\bf 35} (2005) 561).
The model satisfies Einstein's criterion of local causality and is solely based on experimental facts,
comprising the apparatuses used in the experiment and the observation of individual detector clicks.
Our results prove that it is possible to give a particle-only description of
single-photon double-slit experiments.}

\begin{document}

\maketitle

\section{Introduction}
In 1802, Young performed a double-slit experiment with light in order to resolve the question whether light was composed of particles,
confirming Newton's particle picture of light, or rather consisted of waves~\cite{YOUNG}.
His experiment showed that the light emerging from the slits produces a fringe pattern on the screen that is characteristic for interference, discrediting
Newton's corpuscular theory of light~\cite{YOUNG}.
It took about hundred years until Einstein with his explanation of the photoelectric effect in terms of photons, somehow revived Newton's particle
picture of light~\cite{EINS05a}.
In 1924, de Broglie introduced the idea that also matter, not just light, can exhibit wave-like properties~\cite{BROG25}.
This idea has been confirmed in various double-slit experiments with massive objects such as
electrons~\cite{JONS61,MERL76,TONO89,NOEL95}, neutrons~\cite{ZEIL88,RAUC00}, atoms~\cite{CARN91,KEIT91}
and molecules such as $C_{60}$ and $C_{70}$~\cite{ARND99,BREZ02}, all showing interference.
The observation that matter and light exhibit both wave and particle character, depending
on the circumstances under which the experiment is carried out, is reconciled by introducing
the concept of particle-wave duality~\cite{HOME97}.

In most double-slit experiments, the interference pattern is built up by recording individual clicks of the detectors.
In some of these experiments~\cite{MERL76,TONO89,JACQ05} it can be argued that at any time, there is only one object
that travels from the source to the detector.
Under these circumstances, the real challenge is to explain how the detection of individual objects that do not interact with each other
can give rise to the interference patterns that are being observed.

According to Feynman, this phenomenon is ``impossible, absolutely impossible to explain in any classical way and has in it
the heart of quantum mechanics''~\cite{FEYN65}.
Later, Feynman used the double-slit experiment as an example to argue that ``far more fundamental was the discovery that in nature
the laws of combining probabilities were not those of the classical probability theory of Laplace''~\cite{FEYN65b}.
It is known that the latter statement is incorrect as it results from an erroneous application of probability theory~\cite{BALL86,BALL01}.

In this letter, we show that also Feynman's former statement needs to be revised:
We present a simple computer algorithm that reproduces event-by-event,
events being defined as clicks of a detector,
just as in real double-slit experiments, the interference patterns that are usually associated with wave behavior.
We also demonstrate that our event-by-event simulation model reproduces the 
wave mechanical results of a recent single-photon
interference experiment that employs a Fresnel biprism~\cite{JACQ05}.

In our simulation model every essential component of the laboratory experiment such as
the single-photon source, the slit, the Frensel biprism, and detector array
has a counterpart in the algorithm.
The data is analyzed by counting detection events, just as in Ref.~\cite{JACQ05}.
The simulation model is solely based on experimental facts and satisfies Einstein's criterion
of local causality.

In a pictorial description of our simulation model, we may speak about ``photons'' generating
the detection events. However, these so-called photons, as we will call them in the sequel,
are elements of a model or theory for the real laboratory experiment only.
The only experimental facts are the settings of the various apparatuses and the detection events.
What happens in between activating the source and the registration of the detection
events belongs to the domain of imagination.

In the simulation model, the photons have which-way information, never have direct communication with each other
and arrive one by one at a detector.
Although the photons know exactly which route they followed, they
nevertheless build up an interference pattern at the detector, thereby
contradicting the first part of Feynman's first statement, reproduced in the introduction.
With our simulation model, we demonstrate that it is possible to give
a complete description of the double-slit experiment in terms of particles only.

The event-based simulation approach that we describe in this letter
is unconventional in that it does not require knowledge of the wave amplitudes as
obtained by solving the wave mechanical problem.
Instead, the interference patterns are obtained through a simulation of locally causal,
classical dynamical systems.
Our approach provides a common-sense description of the experimental facts without invoking
concepts from quantum theory such as the particle-wave duality~\cite{HOME97}.

In our simulation approach, we adopt the point of view that quantum theory has nothing to say about individual events~\cite{HOME97}.
Therefore, the fact that there exist event-by-event simulation algorithms that reproduce the results of quantum theory
has no direct implications to the foundations of quantum theory:
These algorithms describe the process of generating events at a level of detail
that is outside the scope of what current quantum theory can describe.
The work presented here is not concerned with an interpretation or an extension of quantum theory.

For phenomena that cannot (yet) be described by a deductive theory, it is common practice to use probabilistic models.
Although Kolmogorov's probability theory provides a rigorous framework to formulate such models,
there are ample examples that illustrate how easy it is to make plausible assumptions that create all kinds of paradoxes,
also for every-day problems that have no bearing on quantum theory at all~\cite{GRIM95,TRIB69,JAYN03,BALL03}.
Subtle mistakes such as dropping (some of the essential) conditions,
like in the discussion of the double-slit experiment~\cite{BALL86,BALL01},
mixing up the meaning of physical and statistical independence,
changing one probability space for another during the cause of an argument can give rise to
all kinds of paradoxes~\cite{JAYN89,HESS01,BALL86,BALL03,HESS06}.
It seems that the
interference phenomena that we focus on in this letter
cannot be explained within the framework of Kolmogorov's probability theory~\cite{KHRE01}.
Our simulation approach does not rely on concepts of probability theory:
We strictly stay in the domain of finite-digit arithmetic.

\section{Event-by-event simulation}

In our simulation approach, the photons are regarded as messengers that travel from
a source to a detector. The messenger carries a message that may change as the
messenger encounters another object on its path such as the Fresnel biprism for instance.
The algorithms for updating the messages are designed in such a way that the
collection of many messages, that is the collection of many detection events, reproduces the results of Maxwell's theory.
The key point of these algorithms is that they define classical, dynamical systems that are adaptive.

This letter builds on earlier work~\cite{RAED05d,RAED05b,RAED05c,MICH05,RAED06c,RAED07a,RAED07b,RAED07c,ZHAO08,ZHAO08b} that
demonstrates that it is possible to simulate quantum phenomena on the level of individual events
without invoking concepts of quantum theory or probability theory.
Specifically, we have demonstrated that locally-connected networks of processing units
with a primitive learning capability can simulate event by event,
the single-photon beam splitter and Mach-Zehnder interferometer experiments of Grangier {\sl et al.}~\cite{GRAN86},
Einstein-Podolsky-Rosen experiments with photons~\cite{ASPE82a,ASPE82b,WEIH98},
universal quantum computation~\cite{MICH05,RAED05c},
and Wheeler's delayed choice experiment of Jacques {\sl et al.}~\cite{JACQ07}.

In our earlier work, there was no need to simulate the detection process itself
but, as we argue later, to simulate two-beam interference event by event, it is logically impossible
to reproduce the results of wave theory without introducing a model for the detectors.
We show that the simplest algorithm that accounts for the essential features of a single-event
detector allows us to reconstruct, event by event, the interference patterns that are described by wave theory.
Incorporating this detector model in the simulation models that we reported about in our earlier work 
does not change the conclusions of Refs.~\cite{RAED05d,RAED05b,RAED05c,MICH05,RAED06c,RAED07a,RAED07b,RAED07c,ZHAO08,ZHAO08b}.
In this sense, the detector model adds a new, fully compatible, component to our collection of event-based algorithms.

\begin{figure}[t]
\begin{center}
\onefigure[width=8.5cm]{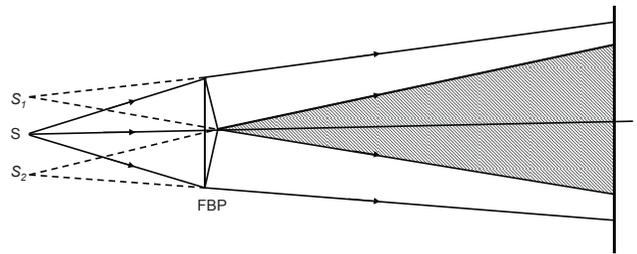}
\caption{Schematic diagram of an interference experiment with a Fresnel biprism (FBP)~\cite{BORN64}.
$S$, $S_1$, $S_2$ denote the point source and its two virtual images, respectively. The grey area is the region
 in which an interference pattern can be observed.}
\label{DS_FBP}
\end{center}
\end{figure}

\begin{figure}[t]
\begin{center}
\onefigure[width=7.5cm]{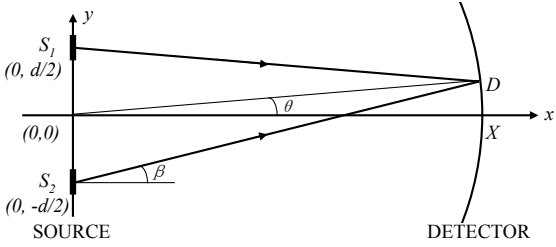}
\caption{Schematic diagram of a double-slit experiment with two sources $S_1$ and $S_2$ of width $a$,
separated by a center-to-center distance $d$.
The sources emit light according to a uniform intensity distribution and with a uniform angular distribution, 
$\beta$ denoting the angle. The light is recorded by detectors $D$ positioned on a semi-circle with radius $X$.
}
\label{doubleslit}
\end{center}
\end{figure}

\begin{figure}[t]
\begin{center}
\onefigure[width=7.5cm]{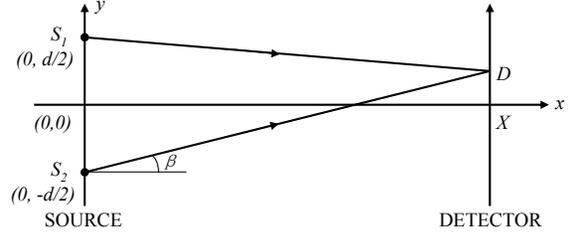}
\caption{Schematic diagram of a two-beam interference experiment with two line sources $S_1$ and $S_2$ having a spatial
Gaussian profile.
The sources are separated by a center-to-center distance $d$ and emit light according to a uniform angular distribution, 
$\beta$ denoting the angle.
The light is detected by detectors $D$ positioned at $(X,y)$.}
\label{simu11}
\end{center}
\end{figure}

\section{Wave theory}

Figure~\ref{DS_FBP} shows a schematic diagram of a two-beam interference experiment with a Fresnel biprism~\cite{BORN64}.
A pencil of light, emitted by the source $S$, is divided by refraction into two pencils~\cite{BORN64}.
Interference can be obtained in the region where both pencils overlap, denoted by the grey area in Fig.~\ref{DS_FBP}.
As a Fresnel biprism consists of two equal prisms with small refraction angle and as the angular aperture of the pencils is small,
we may neglect aberrations~\cite{BORN64}. 
The system consisting of the source $S$ and the Fresnel biprism can then be replaced by
a system with two virtual sources $S_1$ and $S_2$~\cite{BORN64}, see Fig.~\ref{DS_FBP}.
For such a system, it is straightforward to compute the intensity of the wave at the detector screen. 

We consider two cases:
\begin{itemize}
\item{The sources $S_1$ and $S_2$ are slits of width $a$, separated by a center-to-center distance $d$, see Fig.~\ref{doubleslit}.
Then, in the Fraunhofer regime, we have~\cite{BORN64}
\begin{equation}
 I(\theta) = A\left(\frac{\sin\frac{ka\sin\theta}{2}}{\frac{ka\sin\theta}{2}}\right)^2 \cos^2\frac{kd\sin\theta}{2},
\label{eq_slit}
\end{equation}
where $A$ is a constant, $k$ is the wave number, and $\theta$ denotes the angular position of the detector on the circular screen.}
\item{The sources $S_1$ and $S_2$ form a line source with a current distribution given by
\begin{equation}
 J(x,y) = \delta(x)\sum_{s=\pm1}
 e^{-(y-sd/2)^2/2\sigma^2}
 ,
\label{Jy}
\end{equation}
where $\sigma$ is the variance and $d$ denotes the distance between the centre of the two sources.
The intensity of the overlapping pencils reads
\begin{equation}
 I(y) = B\left(\cosh\frac{byd}{\sigma ^2} + \cos\frac{(1-b)kyd}{X} \right)e^{\frac{-b(y^2+d^2/4)}{\sigma ^2}},
\label{eq_Gaussian}
\end{equation}
where $B$ is a constant, $b=k^2\sigma^4/(X^2 + k^2\sigma^4)$, and $(X,y)$ are the coordinates 
of the detector (see Fig.~\ref{simu11}). Closed-form expression Eq.~(\ref{eq_Gaussian}) was obtained
by assuming that $d\ll X$ and $\sigma\ll X$.}
\end{itemize}
From Eqs.~(\ref{eq_slit}) and (\ref{eq_Gaussian}), it directly follows that the intensity distribution on the detector screen
displays fringes that are characteristic for interference.

Results of a time-resolved experiment that is a laboratory realization
of the interference experiment schematically depicted in Fig.~\ref{simu11}
are presented in Ref.~\cite{SAVE02}.

\section{Simulation model}
Looking at Figs.~\ref{doubleslit} and \ref{simu11}, common sense leads to the conclusion that
once we exclude the possibility that there is direct communication between photons,
the fact that we observe
individual events that form an interference pattern can only be due to the presence and
internal operation of the detector (we ignore the possibility that there is no common-sense explanation for this phenomenon).
Therefore, it is worthwhile to consider the detection process in more detail.

In its simplest form, a light detector consists of a material that can be ionized by light.
The electric charges that result from the ionization process are then amplified and
subsequently detected by appropriate electronic circuits.
As usual, the interaction between the incident electromagnetic field ${\mathbf E}$ and
the material takes the form ${\mathbf P}\cdot{\mathbf E}$, where ${\mathbf P}$ is the polarization vector of the material~\cite{BORN64}.
Treating this interaction in first-order perturbation theory, the detection probability is
$P(t)=\int^{t}_{0}\int^{t}_{0}\langle\langle{\mathbf E^{T}(t')}\cdot{\mathbf K(t'-t'')}\cdot{\mathbf E(t'')} \rangle\rangle dt'dt''$
where ${\mathbf K(t'-t'')}$ is a memory kernel that contains information about the material only and
$\langle\langle.\rangle\rangle$ denotes the average with respect to the initial state of the electromagnetic field~\cite{BALL03}.
Very sensitive photon detectors such as photomultipliers and avalanche diodes have an additional feature:
They are trigger devices meaning that the generated signal depends on a threshold.

From these general considerations, it is clear that a minimal model should be able
to account for the memory and the threshold behavior of real photon detectors.
As it is our intention to perform event-based simulations,
the model for the detector should, in addition to the two features mentioned earlier, operate on the basis of individual events.

\subsection{Messenger}
In our simulation approach, we view each photon as a messenger.
Each messenger carries a message, representing its time of flight.
Compatibility with the macroscopic description (Maxwell's theory) demands that the encoding
of the time of flight is modulo a distance which, in Maxwell's theory, is the wavelength of the light.
Thus, the message is conveniently encoded as a two-dimensional unit vector
${\mathbf e}_i=(e_{0,i}, e_{1,i})=(\cos\phi_i, \sin\phi_i)$, where $\phi_i$ is the event-based equivalent of the
phase of the electromagnetic wave and the subscript $i>0$ labels the messages.

\subsection{Source}
A single-photon source is trivially realized in a simulation model in which the photons are viewed as messengers.
We simply generate a message, wait until this message has been processed by the detector, and then generate the next message.
This ensures that there can be no direct communication between the messages, implying that our simulation model (trivially) satisfies Einstein's criterion of local causality.

\subsection{Detector}
A simple model for the detector that accounts for the three features mentioned earlier,
contains an internal vector ${\mathbf p}_i = (p_{0,i},p_{1,i})$ with Euclidean norm less or equal than one.
This vector is updated according to the rule
\begin{equation}
	{\mathbf p}_i = \gamma {\mathbf p}_{i-1} + (1-\gamma) {\mathbf e}_i,
	\label{ruleofLM}
\end{equation}
where $0<\gamma<1$. The machine generates a binary output signal $S_i$ using the threshold function
\begin{equation}
	S_i = \Theta(p^{2}_{0,i}+p^{2}_{1,i}-r_i),
	\label{thresholdofLM}
\end{equation}
where $0\leq r_i <1$.
The total detector count is defined as
\begin{equation}
	N=\sum^{M}_{i=1}S_i,
	\label{N_counts}
\end{equation}
where $M$ is the total number of messages received. Thus, $N$ counts the number of one's generated by the machine.
Although not essential for our simulations, it is convenient to use pseudo-random numbers for $r_i$.
This will mimic the unpredictability of the detector signal. The parameter $\gamma$ controls the precision
with which the message processing machine defined by Eq.~(\ref{ruleofLM}) can represent a sequence of messages with the same
${\mathbf e}_i$ and also controls the pace at which new messages affect the internal state of the machine~\cite{RAED05d}.
The internal vector ${\mathbf p}_i$ and $\gamma$ play the roles of the polarization vector ${\mathbf P}(t)$
and the memory kernel ${\mathbf K(t'-t'')}$, respectively. Notice that the formal solution of Eq.~(\ref{ruleofLM})
has the same mathematical structure as the constitutive equation
${\mathbf P}(t)=\int_0^t \chi(u) {\mathbf E}(t-u) du$ in Maxwell's theory~\cite{BORN64}.
The threshold function of the real detector is implemented through Eq.~(\ref{thresholdofLM}).
A detector screen is just a collection of identical detectors and is modeled as such.
Each detector has a predefined spatial window within which it accepts messages.

It is not easy to study the behavior of the classical, dynamical system defined by
Eqs.~(\ref{ruleofLM})--(\ref{N_counts}) by analytical methods but it is close to trivial
to simulate the model on a computer.

Note that the machine defined by Eq.~(\ref{ruleofLM}) has barely enough memory to store the equivalent of one message.
Thus, the machine derives its power from the way it processes successive messages, not from storing a lot of data.
In particular, the machine does not know about $M$, a quantity that is unknown in real experiments. 

It is self-evident that the detector model defined by Eqs.~(\ref{ruleofLM}) and
(\ref{thresholdofLM}) should not be regarded as a realistic model for say, a photomultiplier.
Our aim is to show that, in the spirit of Occam's razor, this is probably the simplest event-based model
that can reproduce the interference patterns that we usually describe by wave theory.

\begin{figure}[t]
\begin{center}
\onefigure[width=8.5cm]{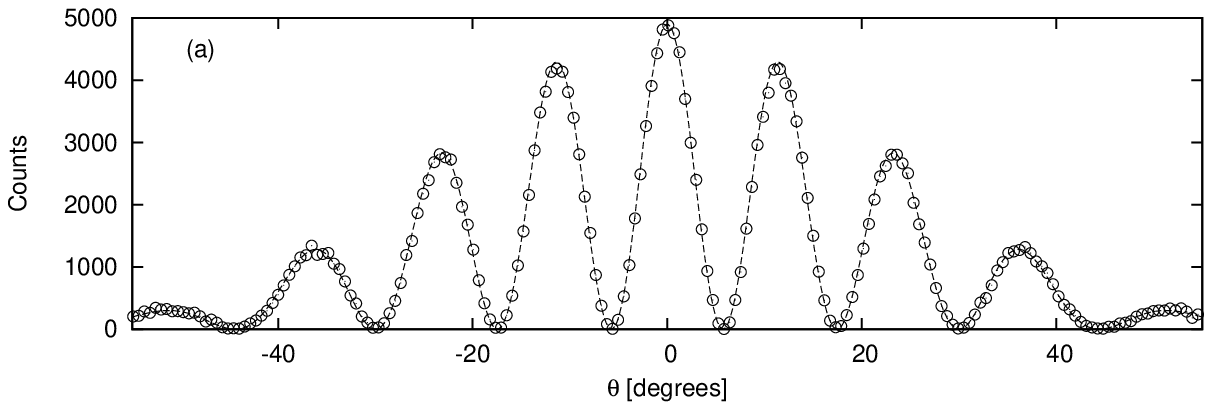}
\onefigure[width=8.5cm]{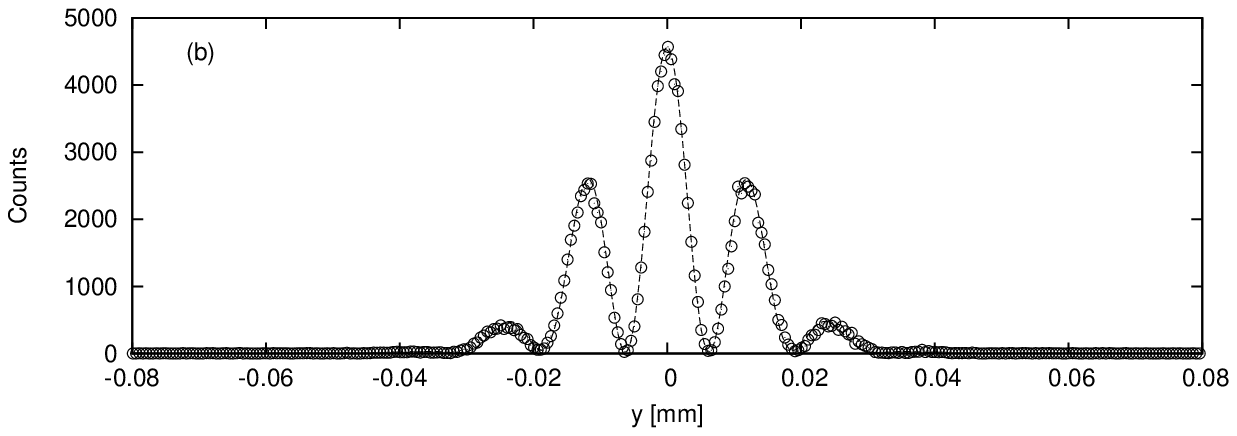}
\caption{Detector counts as a function of the angular (spatial) detector position $\theta$ ($y$) 
as obtained by event-by-event simulations of the interference experiment shown in Fig.~\ref{doubleslit} (Fig.~\ref{simu11}).
The open circles denote the event-based simulation results.
The dashed lines are the results of wave theory (see Eqs.~(\ref{eq_slit}) and (\ref{eq_Gaussian})).
(a) The sources are slits of width $a=\lambda$ ($\lambda=670$ nm in all our simulations), separated by a distance $d=5\lambda$, see Fig.~\ref{doubleslit};
(b) The sources $S_1$ and $S_2$, separated by a distance $d=8\lambda$, emit single photons according 
to a Gaussian distribution with variance $\sigma=\lambda$ and mean $d/2$ and $-d/2$, respectively (see Fig.~\ref{simu11}).
}
\label{simu1}
\end{center}
\end{figure}

\begin{figure}[t]
\begin{center}
\onefigure[width=8.5cm]{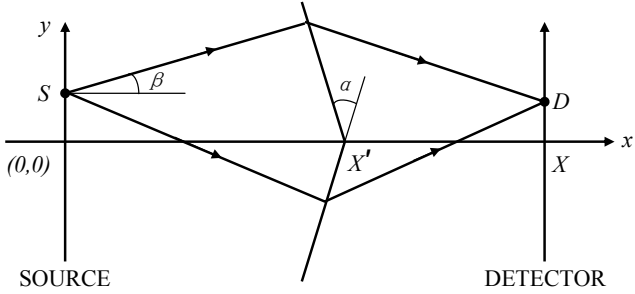}
\caption{Schematic diagram of the simulation setup of a single-photon experiment with a Fresnel biprism.
The apex of the Fresnel biprism with summit angle $\alpha$ is positioned at $(X',0)$.
A line source with a current distribution given by Eq.~(\ref{Jy})
emits single photons from points $S$ and with angles $\beta$ chosen randomly from the interval $[-\alpha/2, \alpha/2]$.
The detectors $D$ positioned at $(X,y)$ count the photons.}
\label{ds_real}
\end{center}
\end{figure}

\begin{figure}[t]
\begin{center}
\includegraphics[width=8.5cm]{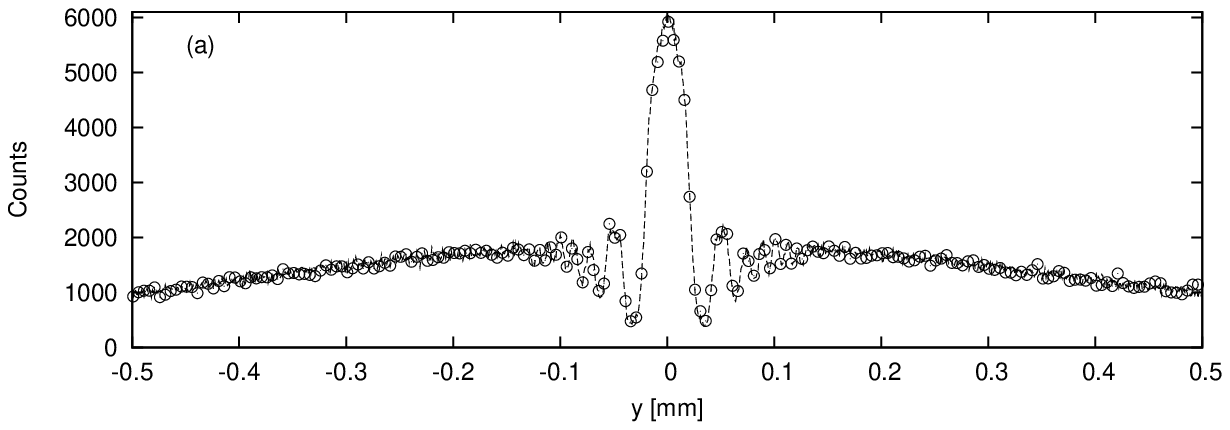}
\includegraphics[width=8.5cm]{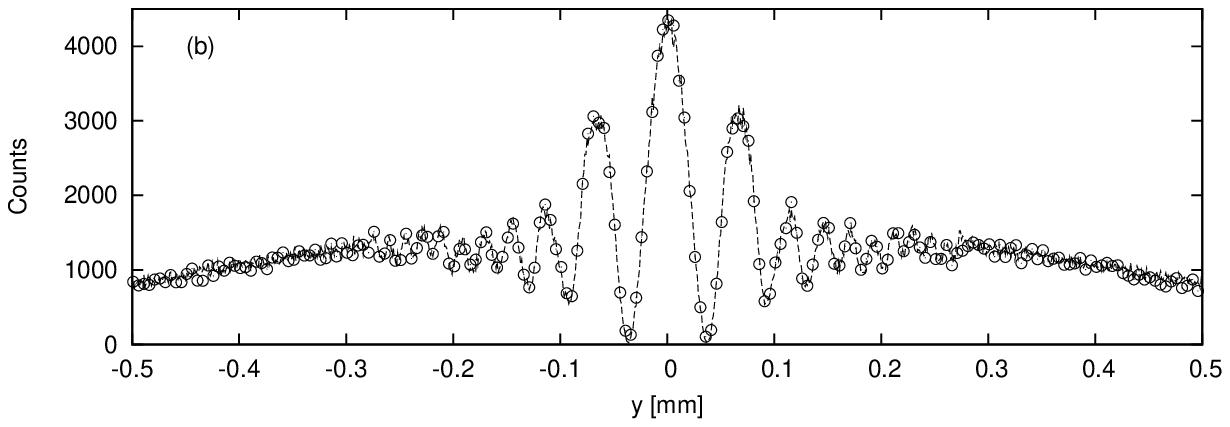}
\includegraphics[width=8.5cm]{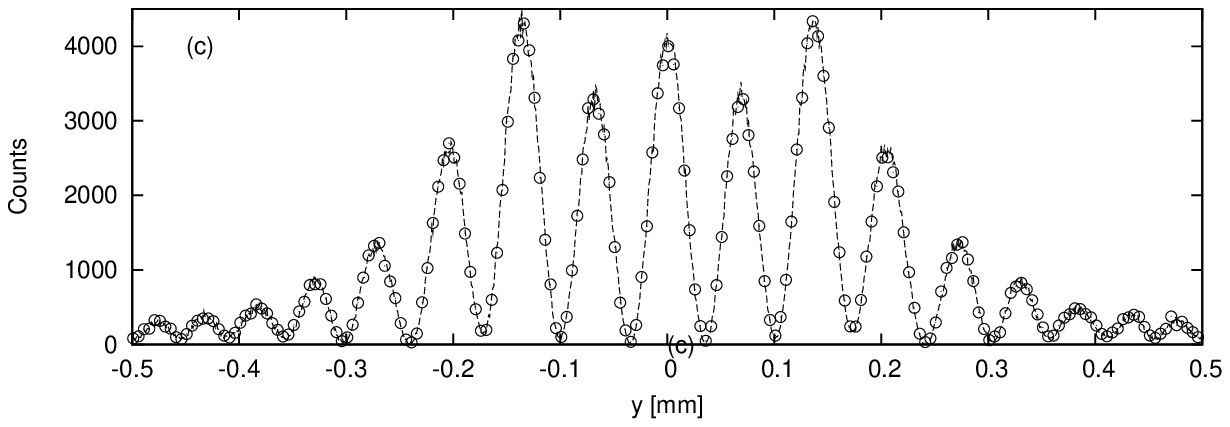}
\caption{Detector counts as a function of the detector position $y$
of the detector array positioned at $X$ (see Fig.~\ref{ds_real}).
The Fresnel biprism has an index of refraction $n=1.5631$ and
a summit angle $\alpha=1^{\circ}$. Its apex is positioned at $(X',0)$ with $X'=45$ mm.
The source emits single photons according to a Gaussian intensity distribution with variance $\sigma=0.531$ mm
and wavelength $\lambda=670$ nm~\cite{JACQ05}.
The open circles denote the event-based simulation results. The dashed lines denote the results as obtained from wave theory.
(a) $X-X'=7$ mm; (b) $X-X'=15$ mm; (c) $X-X'=55$ mm. Thousand detectors where used to record the individual events.}
\label{simu22}
\end{center}
\end{figure}

\section{Simulation results}
First, we show that our event-by-event simulation model reproduces the wave mechanical results
Eq.~(\ref{eq_slit}) of the double-slit experiment.
Second, we simulate a two-beam interference experiment
and demonstrate that the simulation data
agrees with Eq.~(\ref{eq_Gaussian}).
Finally, we present the results for the simulation of the single-photon interference experiment with a Fresnel biprism~\cite{JACQ05},
as depicted in Fig.~\ref{DS_FBP}.

\subsection{Double-slit experiment}

As a first example, we consider sources that are slits of width $a=\lambda$ ($\lambda=670$ nm in all our simulations), separated by a distance $d=5\lambda$,
see Fig.~\ref{doubleslit}.
In Fig.~\ref{simu1}(a), we present the simulation results for a source-detector distance $X=0.05$ mm and for $\gamma=0.999$.
When a messenger (photon) travels from the source at $(0,y)$ to the detector screen positioned at $X$, 
it updates its own time of flight, or equivalently its phase $\phi_i$.
This time of flight is calculated according to geometrical optics~\cite{BORN64}.
As the messenger hits a detector, the detector updates its internal state and decides whether to output a zero or a one.
This process is repeated many times.
The results of wave theory, as given by Eq.~(\ref{eq_slit}), are represented by the dashed lines.
Looking at Fig.~\ref{simu1}(a), it is clear that there is excellent agreement between the event-based simulation and wave theory.

\subsection{Two-beam interference experiment}
As a second example, we assume that the messengers leave either source $S_1$ or $S_2$ from a position $y$ that is
distributed according to a Gaussian distribution with variance $\sigma$ and mean $+d/2$ or $-d/2$, respectively
(see Fig.~\ref{simu11}).
The simulation results for a source-detector distance $X=0.1$ mm and for $\gamma=0.999$ are shown in Fig.~\ref{simu1}(b).
The dashed line is the corresponding result of wave theory, see Eq.~(\ref{eq_Gaussian}).
Also in this case, the agreement between wave theory and the event-by-event simulation is extremely good.

\subsection{Experiment with a Fresnel biprism}

For simplicity, we assume that the source $S$ is located in the Fresnel biprism.
Then, the results do not depend on the dimension of the Fresnel biprism.
Figure~\ref{ds_real} shows the schematic representation of the single-photon interference experiment that we simulate.
Simulations with a Fresnel biprism of finite size yield results that differ
qualitatively only (results not shown). The time of flight of the $i$th message
is calculated according to the rules of geometrical optics~\cite{BORN64}.

In the simulation, the angle of incidence $\beta$ of the photons is selected 
randomly from the interval $[-\alpha /2,\alpha /2]$, where $\alpha$ denotes
the summit angle of the Fresnel biprism.
A collection of representative simulation results for $\gamma=0.999$
is shown in Fig.~\ref{simu22}, together with the results
as obtained from wave theory. Again, we find that there is excellent quantitative agreement between
the event-by-event simulation data and wave theory.
Furthermore, the simulation data presented in Fig.~\ref{simu22} is qualitatively very similar to the results
reported in Ref.~\cite{JACQ05} (compare with Fig.~4(d) and Fig.~5(a)(b) of Ref.~\cite{JACQ05}).

\section{Conclusion}

In this letter, we have demonstrated that it is possible to give a particle-only description for
single-photon interference experiments with a double-slit, two beams, and with a Fresnel biprism. Our 
event-by-event simulation model
\begin{itemize}
\item{reproduces the results from wave theory,}
\item{satisfies Einstein's criterion of local causality,}
\item{provides a pictorial description that is not in conflict with common sense.}
\end{itemize}

We do not exclude that there are other event-by-event algorithms that
reproduce the interference patterns of wave theory.
For instance, in the case of the single-electron experiment with the biprism~\cite{TONO98},
it may suffice to have an adaptive machine handle the electron-biprim interaction without having 
adaptive machines modeling the detectors. We leave this topic for future research.

We hope that our simulation results stimulate the design of new time-resolved single-photon experiments
to test our particle-only model for interference. However, in order to eventually falsify our model,
it would not suffice to simply publish raw experimental data and show
that they do not agree in all details with the model described in this letter.
Indeed, the model that we have presented is far from unique 
and it requires little imagination to see that one can construct similar but more
sophisticated event-based models that can explain interference without making recourse to wave theory.

\section{Acknowledgment}
We would like to thank K. De Raedt, K. Keimpema, and S. Zhao for many helpful discussions.

\bibliographystyle{eplbib}
\bibliography{../../epr}

\end{document}